\documentclass[apj]{emulateapj}
\usepackage{epsfig}

\def\lesssim{\mathrel{\hbox{\rlap{\hbox{%
 \lower4pt\hbox{$\sim$}}}\hbox{$<$}}}}
\def\gtrsim{\mathrel{\hbox{\rlap{\hbox{%
 \lower4pt\hbox{$\sim$}}}\hbox{$>$}}}}

\begin{document}

\title{Hypervelocity intracluster stars ejected by supermassive black hole binaries
}

\author{Kelly Holley-Bockelmann\altaffilmark{1,2}, Steinn Sigurdsson\altaffilmark{1,2}, J. Christopher Mihos\altaffilmark{3}, John J. Feldmeier\altaffilmark{4,5}, Robin Ciardullo\altaffilmark{1} \&
Cameron McBride\altaffilmark{6}}
\altaffiltext{1}{Department of Astronomy, Pennsylvania State University,
kellyhb@gravity.psu.edu}
\altaffiltext{2}{The Center for Gravitational Wave Physics, Pennsylvania State University}
\altaffiltext{3}{Department of Astronomy, Case Western Reserve University}
\altaffiltext{4}{National Optical Astronomy Observatory}
\altaffiltext{5}{NSF Fellow}
\altaffiltext{6}{Department of Astronomy, University of Pittsburgh}

\begin{abstract}

Hypervelocity stars have been recently discovered
in the outskirts of galaxies, such as the unbound star in the Milky Way
halo, or the three anomalously fast intracluster planetary
nebulae (ICPNe) in the Virgo Cluster. 
These may have been ejected by close 3-body interactions with a binary 
supermassive black hole (SMBBH), where a star which passes within the
semimajor axis of the SMBBH can receive enough energy to eject it from the
system. Stars ejected by SMBBHs may form a
significant sub-population with very different
kinematics and mean metallicity than the bulk of the
intracluster stars. The number, kinematics, and orientation
of the ejected stars may constrain the mass ratio,
semimajor axis, and even the orbital plane of the SMBBH. We investigate the evolution of the ejected
debris from a SMBBH within a clumpy and time-dependent cluster 
potential using a high resolution, self-consistent cosmological 
N-body simulation of a galaxy cluster. We show that the 
predicted number and kinematic signature of the fast Virgo ICPNe is consistent with 3-body scattering by a SMBBH with a mass ratio $10:1$ 
at the center of M87. 

\end{abstract}


\section{Introduction}

A significant fraction of the stellar component of a galaxy cluster 
is not confined to any galaxy. This intracluster light (ICL) has 
been identified via planetary nebulae (PNe) (Feldmeier et al. 2004;
Aguerri et al. 2005), Red Giant Branch (RGB)
stars (Durrell et al. 2002), and ultra-deep surface photometry 
(e.g., Feldmeier et al. 2002, 2004; Mihos et al. 2005). 
It is commonly thought that intracluster
stars are stripped from galaxies as the cluster assembles (Merritt 1984), 
via high speed galaxy encounters (Moore et al 1996), interactions with
the cluster potential (Byrd \& Valtonen 1990), or by tidal stripping
in infalling groups (Mihos 2004). 
These processes will generate a
debris field that is highly inhomogeneous, with distinctly non-Gaussian velocities, reflecting an unrelaxed intracluster population (Napolitano et al. 2003).

The continued search for intracluster PNe (ICPNe) has revealed several in the 
Virgo Cluster with radial velocities that are extremely rapid compared 
both to its nearest galaxy and to the background cluster. 
In one pointing $\sim 65$ kpc from M87, 
Arnaboldi et al. (2004) found that of 15 ICPNe radial velocities, 
only 12 were consistent with M87's stellar velocity 
dispersion profile (Romanowsky \& Kochanek 2001) and systemic velocity.
Of the three remaining ICPNe, two had velocities which were offset 
$\sim -900$ km/sec from the systemic velocity of M87 and one had a 
$\sim +1500$ km/sec difference. 

While these ICPNe could be associated with an unrelaxed 
tidally-stripped intracluster stellar population, there is an alternative
explanation. In this paper, we explore the possibility
that the high velocity ICPNe were ejected after interacting with a supermassive binary black hole (SMBBH)
at the center of M87. 
Supermassive black holes (SMBH) are a standard component 
of elliptical galaxies and spiral bulges (e.g., Richstone et al. 1998). 
When galaxies evolve through hierarchical merging, 
each SMBH participates and at some point forms a hard binary.
This implies that many galaxies host
SMBBHs. Perhaps the best evidence for this is the double X-ray bright 
nucleii in 
NGC 6240 (Komossa et al. 2003).

During the hard binary phase, stars that pass between the two SMBHs
siphon energy from the binary's orbit via 3-body scattering, and 
most are ejected with  hypervelocities (Hills 1988; Yu \& Tremaine 2003). 
These SMBBH-driven ejecta may 
constitute a different ICL population 
than stars stripped by dynamical interactions between galaxies; they are 
more likely to be metal rich, and, as we will show, have distinct 
kinematics and spatial structure. This makes the two populations
easy to separate.

We investigate the structure and kinematics of the debris ejected via 
3-body scattering within a high resolution N-body simulation of a cluster 
potential. We model the SMBBH ejection velocities and study
how well the ejected population remains kinematically distinct within 
a clumpy and still evolving cluster potential. We compare the expected 
mass fractions 
of the ICL generated by 3-body interactions with more traditional tidal 
interactions. According to our models, it is 
plausible that the fast ICPNe were ejected after interacting with a 
SMBBH in M87. We discuss how this theory can be tested with
spectroscopic observations.

\section{Generating Hypervelocity Stars}

Galaxy mergers provide an impetus for SMBHs to meet and 
form a bound system. During a galaxy merger, each 
SMBH sinks to the center of the new galaxy potential due to 
dynamical friction, and eventually becomes bound as a SMBBH.
Dynamical friction then continues to shrink the orbit until the binary is hard
(i.e., the separation between each SMBH, $a_{\rm BBH}$, is such that the system tends to lose energy during stellar encounters). Thereafter, further decay is mediated by 3-body 
scattering with the ambient stellar background until the SMBBH becomes so 
close that the orbit can lose energy via gravitational radiation.  
It will thereafter presumably coalesce. The final stages of SMBBH coalescence 
emit so much gravitational radiation that they are extremely likely to be 
detected by the Laser Interferometer Space Antenna (LISA), a planned 
NASA mission to detect gravitational waves, set to launch in the next decade.

When black holes form a hard binary, any star that 
passes nearby takes energy from the system and changes the star's 
velocity (see Yu 2002):
\begin{equation}
{\delta v} \sim { \sqrt{ {3.2 G M_1 M_2} \over { (M_1 + M_2 ) a_{\rm BBH}}}},
\end{equation}
\noindent where $M_1$ and $M_2$ are the black hole masses, and $a_{\rm BBH} = G M_2/ 4 \sigma_0^2$, where $\sigma_0$ is the central velocity dispersion. Since 3-body scattering 
preserves the
z-component of the angular momentum, the ejected debris is confined to a torus
with a major axis aligned with the instantaneous SMBBH orbital plane (Zier \& Biermann 2001)  .
The kinetic energy 
exchange takes place over a timescale:
\begin{equation}
{\delta t} \sim { \sqrt{ {G (M_1+M_2)} \over { {a_{\rm BBH}^3}}}}.
\end{equation}

To model this process in M87, we need to choose appropriate parameters 
for a SMBBH at the galaxy center. This is currently ill-constrained.
If the jet in M87 were a long-lived
structure, we could estimate the masses and separation of the
binary by shape of jet, since a large secondary 
mass could induce periodic structure. However,
the jet's length, along with a reasonable choice for the bulk gas velocity, implies that M87's jet may only be 2-3 Myr old. Thus, it provides 
no real constraint on the component masses. 
We therefore set the total mass of the SMBBH to
$3.3 \times 10^9 M_\odot$ (Harms et al. 1994; Macchetto et al. 1997 ). 
Semi-analytic models of the assembly and growth of SMBHs using an 
extended Press-Schechter formalism indicate that the most probable mass ratio 
for coalescing SMBBHs at low redshift is $M_1/M_2=10$ 
(e.g., Sesana et al. 2004).
Hence, we set the ratio 
of the two SMBHs to be $M_1/M_2=10$. If we assume the 
binary is hard, then the separation is $a_{\rm BBH}= 2.1$ pc, and $\delta v \sim 1350$ km/sec. A full loss cone then implies a SMBBH coalescence timescale of $O(10^{7})$ years, in which case the SMBHs may already have merged (and may even be associated with 
the jet). However, less optimistic assumptions about the loss cone reservoir 
yield a much longer coalescence timescale, suggesting that the 
SMBBH may be observable.

\begin{figure*}
\space\epsfig{file=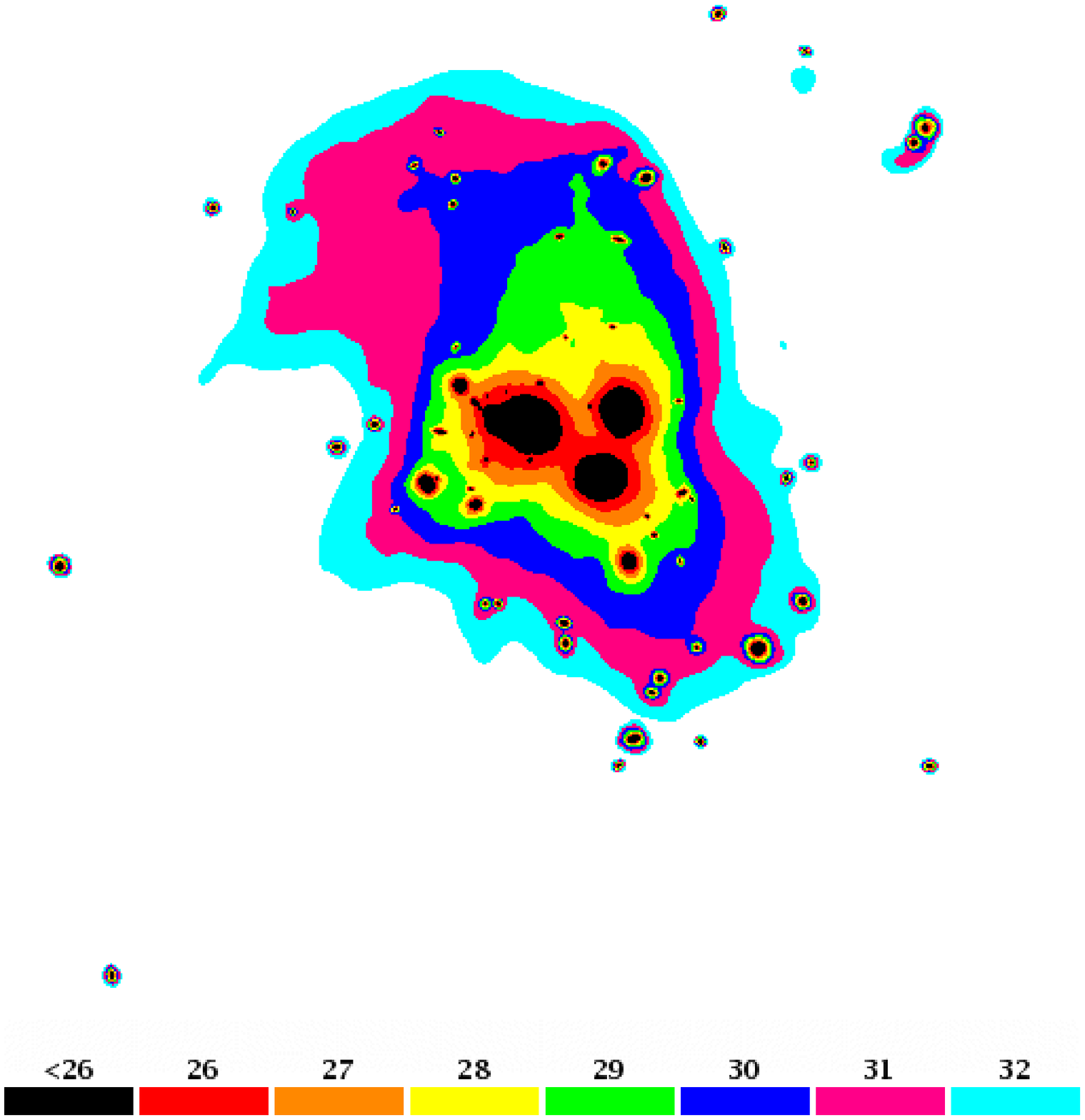, height=2.25in, width=2.25in}\space\space\space\space\epsfig{file=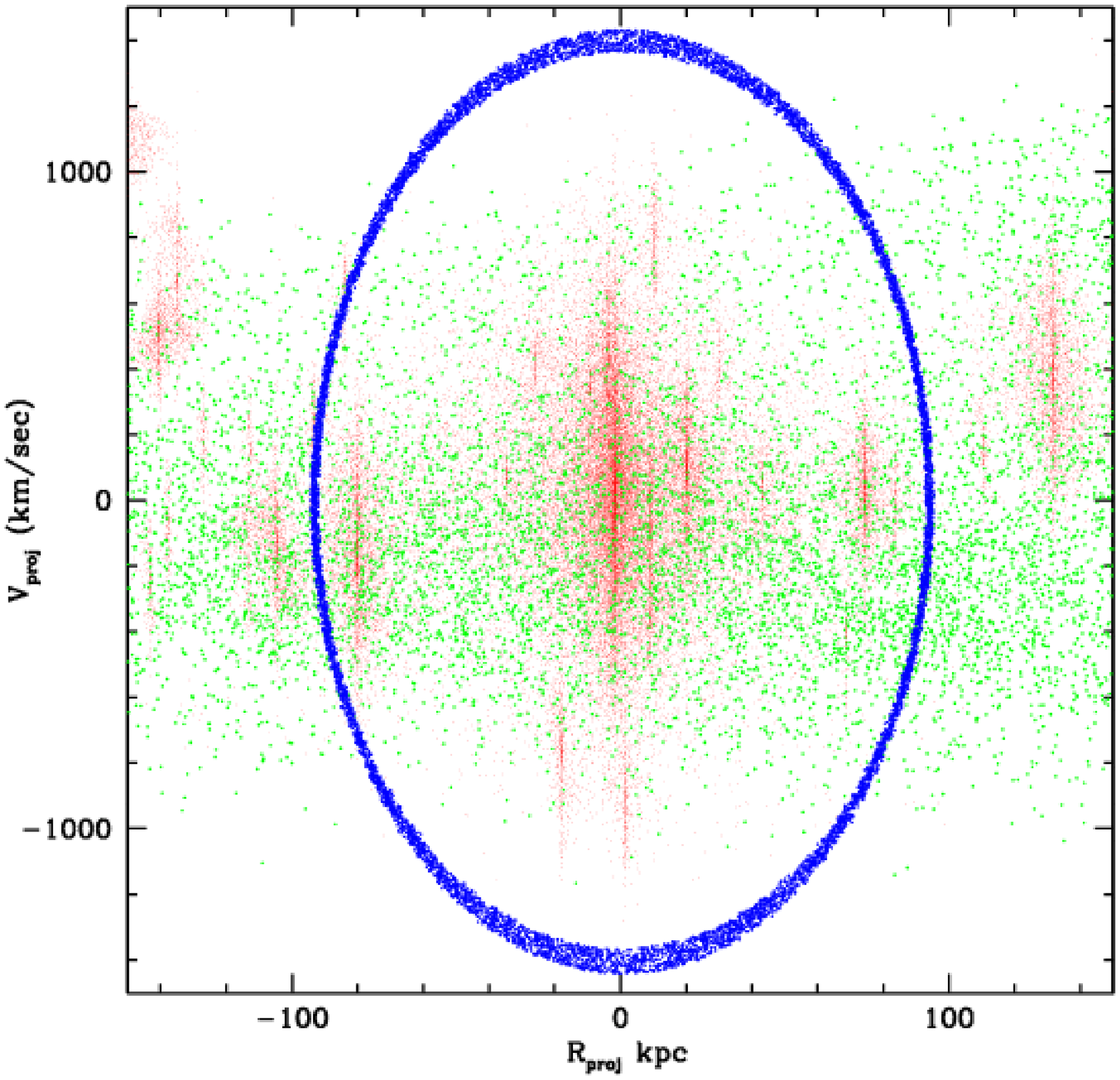, height=2.25in, width=2.25in}\space\space\space\space\epsfig{file=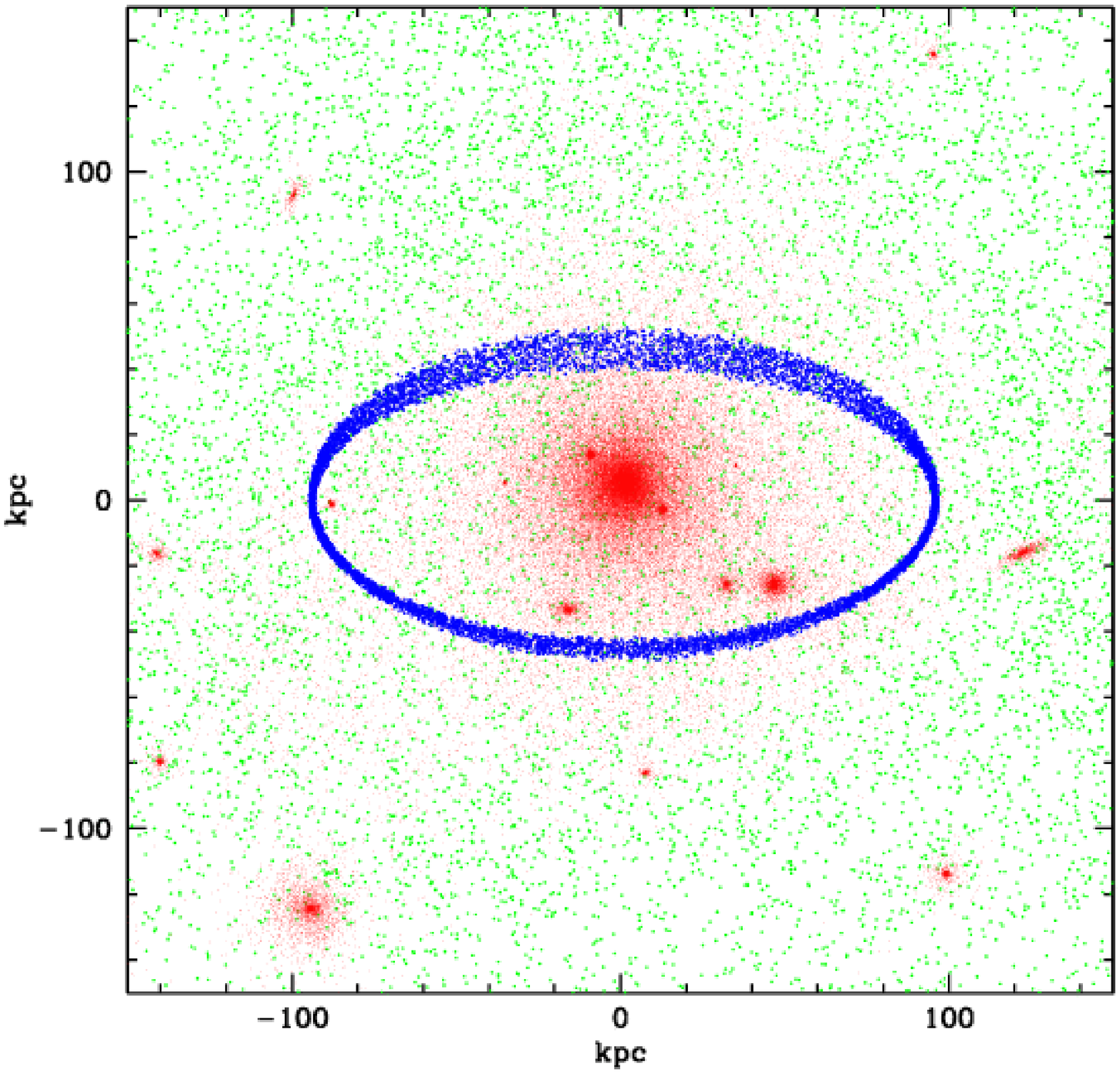, height=2.25in, width=2.25in}
\caption{Left: Surface brightness distribution of the inner 5 Mpc of the 
simulation. Colors represent the surface brightness in the V band. 
Middle: The 
projected position versus line-of-sight velocity for the inner 300 kpc of 
the simulation. The blue points are the particles ejected by the mock SMBBH at the 
center of the most massive galaxy. The ejected debris is clearly distinguishable from 
both the tidally-stripped ICL (green) and the galaxy itself (red). Right: 
Projection of the central 300 kpc around the largest 
galaxy with the observer tiled $50^\circ$ above the orbital plane of the 
SMBBH. The torus thickness is related to the timescale over which the 
debris is ejected. }
\end{figure*}

\section{Simulating Ejecta Kinematics}

To compare the kinematics of the SMBBH-ejected stars to the stellar kinematics
within a complex cluster potential, we use a high-resolution N-body
galaxy cluster simulation as a testbed for our analytic 3-body
ejection scheme (Mihos et al. in prep). To create this simulation, a
50$^3$ Mpc$^3$ $\Lambda$=0.7, $\Omega_M$=0.3 cosmological dark matter
simulation was run from $z$=50 to $z$=0, at which point a collapsed
cluster with mass $\sim 10^{14}$ M$_{\odot}$ was chosen to re-simulate at
higher resolution. At $z$=2, individual halos destined to end up
within the $z$=0 cluster were identified; halos more massive than 10\%
Milky Way mass were excised from the simulation and high resolution
collisionless galaxy models inserted in their place. To do this, we used 
the ``halo
occupancy distribution'' formalism (Berlind \& Weinberg 2002) wherein massive halos are more likely to have
multiple galaxies inserted. Both spiral and elliptical galaxy models
were used. Spirals consisted of an exponential disk
plus a Hernquist (1990) bulge (with bulge:disk ratio of 1:5), while
ellipticals were a pure Hernquist (1990) model. Both galaxy types were
embedded in high resolution isothermal dark  halos which blended
smoothly into the background dark matter distribution. The galaxy
models were scaled in mass to match the halo mass they replaced,
scaled in size by M$^{1/2}$, and scaled in velocity by M$^{1/4}$.  In
total, 121 high resolution galaxy models were inserted into 80 dark
halos; the simulation consisted of approximately 10 million particles,
5.4 million of which represented stars with a gravitational
smoothing length of 0.28 kpc. Initialized at $z$=2, the simulation was
evolved to $z$=0 using the N-body code GADGET (Springel et
al. 2001). Figure 1a shows the simulation at $z$=0  assuming a stellar mass-to-light
ratio of 5.

To compare the characteristics of tidally-released stars with stars ejected 
via 3-body interactions, we must first identify the tidal
ICL. Definitions of ICL abound, based on binding energy, morphology,
or surface brightness.  For our illustrative purposes, we simply
identify particles as ICL if their local density falls below a
threshold, set so that when the cluster is initialized at
$z$=2, the ICL fraction is essentially zero, and at $z$=0 the ICL
particles no longer trace the galaxy morphologies. As a
consistency check, using this criterion we arrive at an ICL fraction
of 13\%, similar to that inferred for Virgo from
observations of ICPNe (Feldmeier et al. 2004).

Even with such a high resolution N-body simulation, no particle would pass 
within $a_{\rm BBH}$ of our M87-analogue nucleus. We therefore developed a 
Monte Carlo approach to mimic 
the interaction between stars and a SMBBH. We pinned
a spherical region with radius $a_{\rm BBH}$ to the center of the most 
massive cluster galaxy (with a mass comparable to M87), and embedded 
the cluster with 8000 particles. We then gave these particles a small 
isotropic velocity dispersion so that they were destined
to pass within $a_{\rm BBH}$ rather quickly. Each particle received the 
velocity kick of equation 1 whenever it passed within $a_{\rm BBH}$ of 
the galaxy center. The direction of this kick was planar, to ensure the 
debris would form a torus. Otherwise, all particles moved 
within the substructure-rich potential. This simulation was followed for 
100 Myr.

Figure 1b and 1c depict the inner 300 kpc of the most massive galaxy.
Here the torus of ejected PNe 
is plainly visible and the radial velocities are typically much faster than 
the ambient ICL. 

\section{Total mass ejected by SMBBHs}

We can estimate the total fraction of ICL ejected from M87 by a SMBBH during 
the hard binary phase:
\begin{equation}
{M_{\rm ej} = {{J} {(M_1 + M_2)}  {\rm ln} \Bigg({{a_{\rm gw}} \over {a_{\rm BBH}}}\Bigg)}},
\end{equation}
\noindent where $J$ is a dimensionless mass ejection rate. For equal-mass 
circular binaries, $J\sim 0.5$ 
(Milosavljevi{\'c} \& Merritt 2001). Quinlan (1996) argues that $J$ drops as $M_1/M_2$ increases, and varies with the orbital speed of the SMBBH. We ignore the variation of $J$ during the binary's evolution,
setting $J=0.3$. The separation where 
gravitational radiation dominates, $a_{\rm gw}$ is:
\begin{equation}
{a_{\rm gw} = { \Bigg({{256} \over {5}} {{G^2 \mu (M_1 + M_2)^2 \sigma_0} \over { c^5 \rho H }}\Bigg)^{1/5}}},
\end{equation}
\noindent where $\rho$ is the central stellar density, $c$ is the speed of light and $H$ is the 
hardening rate, set by restricted 3-body experiments and direct N-body 
simulations. We adopt  $H= 16$ in the estimate that follows (Quinlan 1996), and use
Faber et al. (1997) to set M87's $\rho $ and $\sigma_0$.
Given these constraints, we find that $\sim 6 \times 10^9 M_\odot$ 
should be ejected by a 10:1 SMBBH in M87 during the hard phase.
Although this is a significant amount of mass, it is only about $1.5\%$
of M87's total mass within 10 kpc (Nulsen \& B{\" o}hringer 1995). 
Assuming $M/L=5$, and that M87's V-band luminosity is
$7.7 \times 10^{10} L_\odot$, nearly $2\%$ of M87's light 
could ejected during one 10:1 SMBBH merger.

Scaling this result from M87 to the cluster as a whole requires a 
model for the SMBBH merger history of each Virgo galaxy, along
with SMBH masses
and mass ratios as the cluster assembles. To obtain a rough estimate, however,
we can assume that each Virgo cluster galaxy hosted one 10:1 SMBBH merger in 
its lifetime, and ejected debris during the merger with an efficiency that 
depends on the black hole mass, $M_{\bullet}$. In reality, only a fraction of Virgo cluster
galaxies would host SMBBHs mergers, yet each of these 
would likely have undergone several mergers by $z=0$. Our simple model 
averages over the galaxy ensemble and avoids the problem of identifying 
mergers over a Hubble time. This reduces the problem to simply 
counting the appropriate galaxies and assigning a SMBBH mass to each one.


Using the observed Virgo luminosity function 
(Sandage, Binggeli, \& Tammann 1985), we take a census of the number of
galaxies per absolute blue magnitude, between $-22 < M_B < -12$, as 
well as the variation of morphological type in each magnitude bin.
Since SMBHs are thought to exist in every elliptical and spiral galaxy bulge, 
we include only spirals and ellipticals in our analysis. 
We model all spirals after the Milky Way, with the same mass, $\sigma_0$, 
and bulge to disk ratio of 1:3; our elliptical galaxies are modeled using 
Fundamental Plane parameters (Dressler et al. 1987). Naturally, spiral galaxies
are more varied than the Milky Way, but as we will see, the choice of our
spiral model matters very little for this exercise. 

Now that we have the number of 
SMBBH-embedded galaxies, as well as $\sigma_0$ for each, we determine the
SMBBH mass in each galaxy using the  $M_\bullet-\sigma$ relation 
(Gebhardt et al 2000; Ferrarese \& Merritt 2000).
Unfortunately, there is some disagreement on the precise
slope of the relation, so we adopt $M_\bullet \propto \sigma_0^{4}$ (e.g., Tremaine et al. 2004). 
Since we have the SMBBH mass and $\sigma_0$, we also have $a_{\rm BBH}$ and
$a_{\rm gw}$, and hence $M_{\rm ej}$ for each galaxy.


Folding in these various relationships and summing over the luminosity function, we obtain a total SMBBH-ejecta luminosity  $\sim 2\%$ of the total Virgo Cluster luminosity. SMBBHs in low luminosity ellipticals
contribute slightly more per galaxy to the ICL ($3\%$) than the 
brightest ellipticals, but most SMBBH ejections occur in the brightest 
magnitude bins, which are occupied by large ellipticals like M87. 
The contribution from spirals is negligible ($0.01\%$).  Of course, 
the fraction of light from galaxy interactions is more substantial; N-body 
simulations suggest $10-50\%$ of the luminosity of 
the Virgo Cluster could come from intracluster stars stripped during galaxy 
interactions (Napolitano et al. 2003). However, this study indicates that as much as $20\%$ of
the ICL could come from the more exotic SMBBH origin. We caution that since this 
estimate clearly hinges on $M_{\bullet}$, varying the 
$M_\bullet-\sigma$ relation can change $M_{\rm ej}$ (and hence, the fraction of SMBBH-generated ICL) significantly.



\section{Implications}

Intracluster stars may be released via two very distinct mechanisms: 
tidal stripping of the outskirts of galaxies, and 3-body interactions with 
a SMBBH deep within a galaxy 
potential well. 
Stars ejected by a SMBBH are kinematically separable
from the ambient cluster kinematics and also preserve an easily discernible
shape, even in a complex and evolving cluster potential.

A thin torus is the $prompt$ ejection signal from a single SMBBH as 
it forms a hard binary. Hence, observing
a torus of ICPNe can constrain both the recent existence of a SMBBH and its 
orbital plane without sub-arcsecond observations.
For example, given a full loss cone, it takes $\sim 10^7$ years for a 
$10:1$ SMBBH to evolve from a hard binary to one that decays via gravitational 
radiation. During that time, $a_{\rm BBH}$ shrinks, lowering the stellar 
interaction cross-section but strongly increasing the few ejected 
velocities. So, if the hypervelocity ICPNe observed are indeed caused 
by an interaction with a SMBBH, the last major galaxy merger was complete less 
than $10^7$ years ago. With a $10:1$ galaxy merger timescale of $O(10^8)$ 
years, there should be some fine structure (Schweizer \& Seitzer 1992) 
within M87 associated with the event (and there may be: see Weil et al. 1997). 

The current consensus is that most ellipticals have assembled by merging 
(e.g., van Dokkum 2005). Therefore, in a cluster environment, debris may 
originate from several galaxies over many epochs, which makes the ejection 
signature more complex. To determine the aggregate behavior of SMBBH-driven ICL
we plan to investigate the long-term time evolution of the ejecta from 
SMBBHs as galaxies assemble within a cluster environment.

The ICL fraction generated by SMBBHs ($4-20\%$) is consistent with 
the high velocity ICPNe fraction in Virgo, $\sim 7\%$. 
Hence, the three ICPNe outliers could all be hypervelocity ejections. 
As a further reality check, given the predicted $O(10^3)$ km/sec velocity 
kick, it takes $O(10^7)$ years to travel $65$ kpc from M87. Since the 
lifetime of a $2 M_\odot$ star is $O(10^8)$ years,  
it is plausible that a main sequence star could plunge from the inner kpc, 
get ejected by the SMBBH, and evolve into a PNe within the allotted time.
The tidal disruption radius of such a star is $O(10^{-5})$ pc, 
so it  may pass within $a_{\rm BBH}$ intact. 

There is another way to determine the importance of SMBBH 
ejections.
Since this mechanism works deep within the galaxy potential,
the ejected stars should retain the metallicity of the elliptical galaxy 
center.
Observations indicate that bright ellipticals have a mean abundance in the 
range $0<\big[ Z/H\big]<0.4$ and a metallicity gradient of
$\Delta {\rm log} Z/\Delta {\rm log} R\sim-0.25$ (Henry \& Worthey 1999).
This implies that SMBBH
ejection candidates will typically have a much higher metallicity than 
the tidally generated ICL, which usually 
originate in the outskirts of a galaxy. It would be ideal if this 
metallicity divide were clear-cut, but some scatter in the
abundances of both types of ICL is expected.
Recall that, by far, most of the SMBBH ejecta originate 
from the center of the largest ellipticals, and should therefore 
exhibit high metallicities. This high 
metallicity sample is contaminated, though, with a few 
low metallicity SMBBH 
ejections that originate from faintest ellipticals (Bender et al. 1993).
Strong galaxy interactions, too, can occasionally dredge up metal-rich material from the inner regions of galaxies (Hibbard \& Mihos
1995; Mihos 2006 in prep), after which the cluster potential can
easily strip it into the ICL. Finally, since the 
scatter in $\big[ Z/H \big]$ is large within a galaxy,  metal poor stars 
exist even within a metal rich bulge (e.g., Jacoby \& Ciardullo 1999). 
Despite these complications, since the {\it mean} metallicities of each 
population differ, multi-night spectroscopic observations with an 
8-meter telescope (or better yet, short observations with a 30-meter class 
telescope) can divide the population 
statistically.

The implications for LISA are profound.
If a star passes within 100 Schwarzschild radii
of either SMBH, it can generate a burst of gravitational
radiation with a signal-to-noise ratio of $O(10^3)$ in the LISA waveband, high enough to be 
observed in Virgo at a rate of $10 $ per year (Rubbo, Holley-Bockelmann, \& Finn, in prep.).  
In the future, we may use LISA observations in concert with hypervelocity PNe
 to confirm and further constrain the masses, mass 
ratios, separation, and even the SMBBH orbital plane within galactic nucleii.

\acknowledgments

KHB is supported by the Center for 
Gravitational Wave Physics, which is funded by the NSF under the cooperative agreement PHY 01-14375. 
This work was completed with the support of a grant from NASA, ATP NNG04GU99G.
JCM is supported by the NSF through grant AST-9876143 and by a Research 
Corporation Cottrell Scholarship. JF is supported through grant 
AST-0302030 (NSF).


\clearpage

\end{document}